\documentclass[preprint2]{aastex}

\def\kms{\rm km\ s^{-1}}

\def\kpch{\mbox{$h^{-1}$kpc}}

\def\mpch{\mbox{$h^{-1}$Mpc}}

\def\hmpc{\mbox{$h^{-1}$Mpc}}
\def\Mpch{\mbox{$h^{-1}$Mpc}}

\def\Msunh{\mbox{$h^{-1}$M$_\odot$}}

\def\LCDM{{$\Lambda$CDM}}
\def\m3{MARK III}

\def\be{\begin{equation}}
\def\ee{\end{equation}}

\def\mathnew{\mathsurround=0pt}
\def\simov#1#2{\lower .5pt\vbox{\baselineskip0pt\lineskip-.5pt\ialign{$\mathnew#1\hfil##\hfil$\crcr#2\crcr\sim\crcr}}}

\def\'#1{\ifx#1i{\accent"13\i}\else{\accent"13#1}\fi}

\shorttitle{Observational signatures of IGM}
\shortauthors{Kravtsov et al.}

\begin{document}
\title{Constrained Simulations of the Real Universe: II. Observational Signatures of
Intergalactic Gas in the Local Supercluster Region}

\author{Andrey V. Kravtsov\footnote{Hubble Fellow; {\em Current address: Department of Astronomy \& Astrophysics, University of Chicago, 5640 S. Ellis Ave., Chicago, IL 60637, US }}}
\affil{Department of Astronomy, The Ohio State University, 140 West 18th Ave., Columbus, OH 43210-1173}
\author{Anatoly Klypin}
\affil{Astronomy Department, New Mexico State University, Las Cruces, NM 88003-0001}
\author{Yehuda Hoffman}
\affil{The Racah Inst. of Physics, Hebrew University, Jerusalem 91904, Israel}

\begin{abstract}
  We present results of gasdynamics$+N$-body {\em constrained}
cosmological simulations of the Local Supercluster region (LSC; about
$30\mpch$ around the Virgo cluster), which closely mimic the real
Universe within 100~Mpc by imposing constraints from \m3 catalog of
galaxy peculiar velocities.  The simulations are used to study the
properties and possible observational signatures of intergalactic
medium in the LSC region.  We find that, in agreement with previous
unconstrained simulations, $\approx 30\%$ of the gas in this region is
in the warm/hot phase at $T\sim 10^5-10^7$~K, and $\approx 40\%$ in
the diffuse phase at $T< 10^5$~K in low-density regions.  The X-ray
emission from the warm/hot gas may represent a small ($\sim 5-10\%$)
but important contribution to the X-ray background observed by the
ROSAT All-Sky Survey at energies around 1 keV. The best prospects for
detection of the warm/hot intergalactic medium of the LSC located in
filaments and in the vicinity of virialized regions of groups and
clusters are through absorption in resonant lines of OVII and OVIII in
soft X-rays and in the OVI doublet in UV. If intergalactic gas in
filaments ($\rho/\langle\rho\rangle\sim 1-10$) is enriched to typical
metallicities of $\gtrsim 0.05$, the column densities of OVI, OVII,
and OVIII along a random line of sight near the North Galactic Pole,
especially near the supergalactic plane, have a significant
probability to be in the range detectable by the current ({\sl FUSE},
{\sl XMM}) and future ({\sl Constellation-X}) instruments.
\end{abstract}


\section{Introduction}
\label{sec:intro}

At the present epoch baryons in stars and galaxies constitute only a
small fraction of the total baryon density predicted by the Big Bang
nucleosynthesis model \citep*{fukugita_etal98}.  This is in agreement
with the Cold Dark Matter (CDM) models of structure formation which
predict that a large fraction of baryons at $z\approx 0$ is located in
filaments and intergalactic medium of groups and clusters, where it is
shock-heated to temperatures of $T\sim 10^5-10^7$~K
\citep{cen_ostriker99,dave_etal99}.

The detection of the warm/hot intergalactic medium (WHIM) and studies
of its properties represent a challenge.  The integrated soft X-ray
emission of this gas may contribute significantly to the observed
extragalactic X-ray background
\citep[XRB;
e.g.,][]{croft_etal01,phillips_etal01,voit_bryan01}.  The best
prospects for direct detection are, however, through absorption or
emission in the lines of ionic species of heavy elements such as
oxygen, which requires very sensitive X-ray and UV spectroscopy of
bright background sources
\citep{shapiro_bahcall80,basko_etal81,aldcroft_etal94,hellsten_etal98,
perna_loeb98,fang_canizares00}.  The recent launches of new generation
X-ray ({\sl Chandra\/} and {\sl XMM-Newton}) and UV ({\sl FUSE})
satellites, as well as access to the UV wavelengths with the {\sl
Hubble Space Telescope (HST)}, led to first detections of absorption
in the resonance OVI doublet in UV at both low
\citep{tripp_savage00,tripp_etal00}
and intermediate
\citep{reimers_etal01} redshifts.  The first attempt to
detect resonance line absorption of heavy metals in soft X-rays with
{\sl Chandra} by \citet{fang_etal01} yielded only an upper limit, but
more sensitive studies are currently underway.

Spectroscopic studies of the warm/hot medium are very promising for
understanding properties and evolution of this important baryonic
component. An important piece of information is the state of the WHIM
gas in the nearby Universe at $z=0$.  However, the absorption at very
low redshifts should be due to only a few large-scale structures and
one would therefore expect model predictions to be rather uncertain
due to the cosmic variance. Moreover, we know that the Local Group is
surrounded by the large-scale filamentary structures in the density
field such as the Local Supercluster, Perseus-Pisces supercluster, the
Great Attractor, etc.  They are expected to contain warm/hot gas and
could serve as targets for its detection. It would be interesting
therefore to explore observational signatures of this gas and make
theoretical predictions on the sky coverage and column density
distributions of gas in these structures.

Fortunately, this problem can be addressed by using {\em constrained
simulations}
\citep[][hereafter Paper I]{klypin_etal:lsc1}, i.e.,
cosmological simulations with initial conditions constrained to
reproduce the observed large-scale structures in the nearby Universe
\citep[see
also][]{kolatt_etal96,bistolas_hoffman98,wh99}.  The initial
conditions are generated within the framework of a flat CDM cosmogony
by means of constrained realizations of Gaussian fields, where the
constraints are set by the \m3\ survey of peculiar velocities.  The
setup of these realizations is such that that the structure on scales
larger than a few Mpc is tightly constrained by the data, while
smaller scales essentially constitute a random realization of the
assumed primordial perturbation field.

In the study presented here, we used the method of
\citet{klypin_etal:lsc1} to set
up and run constrained $N$-body$+$gasdynamics simulations of the Local
Supercluster  region (LSC;  a roughly  spherical region  of $R\lesssim
30\mpch$  centered on the  Virgo cluster)  in {\LCDM}  cosmology.  The
simulation  followed   gasdynamics  of  the   baryonic  component  and
collisionless  dynamics  of  dark   matter  particles  using  a  newly
developed eulerian adaptive mesh refinement code (see
\S~\ref{sec:code}). Gas cooling (except for adiabatic expansion) was
not included. The main results of our study depend on the physics of
shock-heated ($T\gtrsim 10^5$~K) gas in the relatively low density
($\rho/\langle\rho\rangle\sim 1-10$) regions.  The cooling time of
such gas is $t_{\rm cool}\approx 3.1\times 10^{11}\ {\rm yrs\ K^{-1}}
\Lambda_{-23}(1+\delta)^{-1}T_5$, where we assumed $\Omega_b =
0.02h^{-2}$ and complete ionization of gas, $\Lambda_{-23}$ is cooling
rate in units of $10^{-23} {\rm cm^3\ erg\ s^{-1}}$, $\delta$ is gas
overdensity, and $T_5$ is gas temperature in $10^5$~K. The cooling
time is thus longer than the Hubble time in our regions of interest
and effects of cooling can be neglected. The cooling time can be short
in the cluster cores where gas cooling can result in a cooling
flow. Such regions, however, occupy a very small fraction of the
volume and will not affect the main conclusions of the paper about the
probability for the WHIM detection via metal absorption.  The cooling
will affect our results on the X-ray fluxes from regions in the
cluster cores; our fluxes should be viewed as an upper limit on the
possible emission.

The use of observational constraints dictates that the simulations are
initialized and evolved in the supergalactic coordinates used to map
the actual universe.  This enables us to construct sky maps of the
desired quantities (e.g., X-ray surface brightness, or column
densities) corresponding to real objects and actual locations in the
real universe.  The large angular extent of the LSC and its favorable
orientation on the sky\footnote{The concentration of nearby galaxies
along the supergalactic plane extends over more than $100^{\circ}$.
The plane is almost perpendicular to the galactic plane; its center,
the Virgo cluster, is close to the North Galactic Pole.}  imply that a
large fraction of observational lines-of-sight in the northern
galactic hemisphere pass through the regions of high gas density.  We
analyze the distribution and properties of gas in the LSC region
focusing on its possible observational signatures such as absorption
and emission by OVI, OVII, and OVIII, soft X-ray emission, and
Sunyaev-Zeldovich effect. 

The paper is organized as follows. The numerical code, initial
conditions and simulations are described in \S~\ref{sec:code} and
\S~\ref{sec:sims}, respectively.  The geometry of density
field, properties of gas in the simulation, sky maps of various
quantities, and observational signatures of the gas are discussed in
\S~\ref{sec:results}.  We conclude the paper with a brief discussion
of the results and our conclusions in \S~\ref{sec:discussion}.

\section{Numerical code}
\label{sec:code}

The adaptive refinement tree (ART) code \citep{kkk97,kravtsov99}
achieves high spatial resolution by adaptively refining regions of
interest using an automated refinement algorithm.  The $N$-body part
of the code was used extensively to run high-resolution
dissipationless cosmological simulations
\citep[e.g.,][ and references therein]{kravtsov99}. 
Due to the use of eulerian mesh hierarchy in the ART algorithm, the
inclusion of eulerian gasdynamics is a natural extension. The adaptive
mesh refinement methods for numerical hydrodynamics enjoy widespread
popularity in the physics and engineering communities and are now also
gaining popularity in cosmology and astrophysics
\citep[e.g.,][]{truelove_etal98,norman_bryan99,khokhlov_etal99,abel_etal00,plewa_mueller01,refregier_teyssier01,ricker_etal01}.
The adaptive mesh refinement is particularly attractive for cosmology
because interesting regions in cosmological simulations usually occupy
only a small fraction of the computational volume and thus can be
refined with relatively small number of mesh cells.

Here we will briefly describe the gasdynamics part of the algorithm,
detailed account of which will be presented elsewhere \citep[see
also][]{kravtsov99}.  The equations of gasdynamics and particle motion
are integrated in ``supercomoving'' variables
\citep{shandarin80,martel_shapiro98}. These variables are
remarkable as their use almost completely (completely for ideal gas
with $\gamma=5/3$) eliminates explicit dependence on cosmology in the
model equations. The equations can therefore be integrated using
standard gravitation and gasdynamics solvers with no need for
additional coefficients and corrections.

A simulation is usually started with a uniform grid covering the
entire computational domain.  As the universe evolves and expands,
additional resolution is required in and around dense objects.  The
ART code uses mesh refinement to increase spatial resolution in
regions where such increase is needed.  The refinements are recursive:
the refined regions can also be refined, each subsequent refinement
having half of the previous level's cell size. This creates a
hierarchy of refinement meshes of different resolution, size, and
geometry covering regions of interest.  Because each individual cubic
cell can be refined, the shape of the refinement mesh can be arbitrary
and effectively match the geometry of the region of interest.  This
strategy is particularly well suited for simulations of a selected
region within a large computational box, as in the constrained
simulations presented below.

\begin{figure*}[p]
\includegraphics[width=\textwidth]{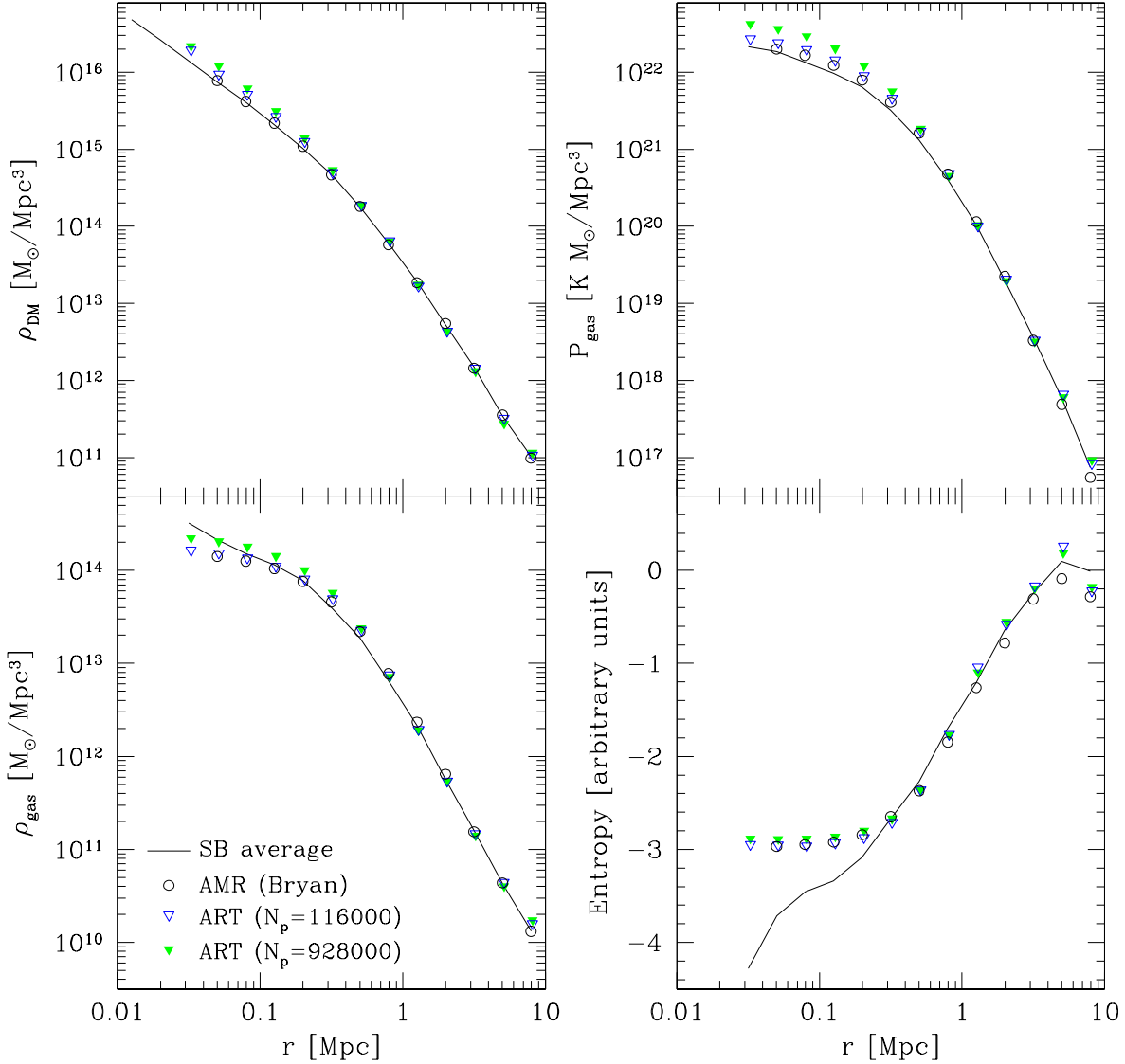}
\epsscale{1.01}
\caption{\small Results of the $N$-body$+$gasdynamics simulations
of the ``Santa Barbara'' cluster using the ART code. The panels show
various spherically averaged profiles; counter clockwise from the
upper left: dark matter density; gas density; gas pressure; gas
entropy.  The ART profiles ({\em triangles}) are compared to the
corresponding mean profiles from the SB cluster comparison project
(Frenk et al.  1999) shown by the {\em solid line}, which represent
average over 12 (13 for the $\rho_{DM}$ profile) simulations performed
using different cosmological codes. The {\em open circles} show
profiles from the adaptive mesh refinement (AMR) simulation by Greg
Bryan performed as part of the SB project. The {\em open triangles}
show profiles for a simulation with the dark matter particle of
$1\times 10^9{\ }{\rm M_{\odot}}$ ($\approx 116,000$ particles within
the cluster virial radius), while {\em solid triangles} show results
of a simulation with eight times better mass resolution ($\approx
928,000$ particles within the virial radius). The profiles from the
ART simulations are shown at scales larger then 4 formal resolutions.
}\label{fig:sb4}
\end{figure*}

During a simulation, spatial refinement is accompanied by temporal
refinement.  Namely, each level of refinement, $l$, is integrated with
its own time step $\Delta \tau_l=\Delta \tau_0/2^l$, where $\Delta
\tau_0$ is the global time step of the zeroth refinement level in the
code's time units (the code uses time variable defined as follows:
$d\tau \equiv a^{-2}dt/t_0$, where $t$ is the physical time, $a$ is
expansion factor, and $t_0$ is an arbitrary time constant).  This
variable time stepping is very important for accuracy of the results.
As the force resolution increases, more steps are needed to integrate
the trajectories accurately. To increase mass resolution in certain
regions the code uses particles of different masses.  In the
simulations presented in this paper this ability was used to achieve
high mass resolution inside a region centered around the Virgo cluster
(see below).

The main features of the gasdynamics implementation follow closely the
algorithm of \citet{khokhlov98}. A second-order Godunov-type solver
\citep{collela_glaz85} is used to compute numerical
fluxes of gas variables through each cell interface, with ``left'' and
``right'' states estimated using linear piecewise linear
reconstruction \citep{vanleer79}. The Riemann solver was written and kindly
provided to us by A.M.~Khokhlov and has good shock-capturing
characteristics (Khokhlov 1998; Kravtsov 1999) which is desirable for
cosmological applications.  The gasdynamics is coupled to the dynamics
of dark matter through the common potential, which is used to compute
accelerations for both the DM particles and the gas. The code employs
the same data structures and similar refinement strategy as the
previous $N$-body versions of the code.  However, in addition to the
DM density criteria it also allows for additional refinement criteria
based on the local density of gas, shock, sharp gradient indicators.
The refinement criteria can be combined with different weights
allowing for a flexible refinement strategy that can be tuned to the
need of each particular simulation.

Extensive tests of the $N$-body code and comparisons with other
numerical $N$-body codes can be found in \citet{kravtsov99} and
\citet{knebe_etal00}. To compare performance of
the $N$-body$+$gasdynamics code to other existing cosmological
gasdynamics codes in a realistic cosmological setting, we have
simulated the cluster used in the ``Santa Barbara (SB) cluster
comparison project'' \citep{frenk_etal:sb}.  This project compared
simulations of the same Coma-size cluster performed using 12 different
gasdynamics codes.  Figure~\ref{fig:sb4} shows spherically averaged
profiles of dark matter density, gas density, gas pressure, and gas
entropy for the ART simulations with two different mass resolutions
for the dark matter.  The ART results are compared to the average
profiles from the SB project and to the profile of the Adaptive Mesh
Refinement (AMR) simulation\footnote{Note that this is entirely
independent AMR code that uses different solvers for both gas and dark
matter and different refinement algorithm.} by G.L.~Bryan performed as
part of the SB project.  The refinements in the ART runs were based on
the local gas density, with the threshold density for refinement level
$l$ equal to $4\times \langle m_{\rm gas}\rangle\Delta x_0^{-3l}$,
where $\langle m_{\rm gas}\rangle $ and $x_0$ are average gas mass per
and size of zeroth-level grid cell, respectively.  The simulations
used $128^3$ uniform grid and six levels of refinement, which
corresponds to the smallest cell size of $\approx 8$ kpc in the 64 Mpc
computational box which is similar to the formal resolution of the AMR
simulation.

The figure shows very good general agreement between the ART and
Bryan's AMR profiles, especially for the lower-resolution ART run
which has the same mass resolution as the AMR run.  The higher mass
resolution run produces a somewhat steeper dark matter density profile
in the innermost regions. This is in agreement with recent convergence
studies for dark matter density profiles \citep{ghigna_etal00,kkbp01}.
It is interesting to note that the steepening of the DM profile
affects gas profiles, resulting in a somewhat higher central gas
density (and smaller core radius) and higher central temperature and
pressure. The entropy profiles of the two eulerian mesh refinement
simulations agree very well and show presence of a well resolved core.
There is a clear and alarming mismatch of these profiles and the
average SB profile which is dominated by the SPH simulations at small
scales.  The origin of the mismatch is unclear, although the fact that
the two different eulerian codes agree well probably indicates that it
is due to less accurate treatment of shocks in the SPH codes. We are
currently working on more detailed comparisons of entropy evolution in
ART and SPH codes.

\section{Constrained simulations}
\label{sec:sims}

The goal of this paper is to perform numerical simulations that match
the observed local universe as well as possible.  Namely, we are
interested in reproducing the observed structures: the Virgo cluster,
the Local Supercluster (LSC) and the Local Group (LG), in the
approximately correct locations and embedded within the observed
large-scale configuration dominated by the Great Attractor and the
Perseus-Pisces supercluster.

The setup details of the constrained initial conditions used for our
simulations are presented in \citep{klypin_etal:lsc1}.  Initial
conditions are constructed using constrained realizations based on the
\m3\ catalog of radial peculiar velocities and assuming the {\LCDM}
model with the following parameters: $\Omega_0=0.3$,
$\Omega_{\Lambda}=0.7$, $\sigma_8 =0.9$, $\Omega_{baryon}=0.03$,
$h=0.7$. Constrained realizations of Gaussian fields are generated by
sampling the random residual from the mean (Wiener filter) field given
the data and the assumed model \citep{hoffman_ribak91}.  The nature of
the data and the {\LCDM} model implies that scales larger than a few
Mpc are tightly constrained by the data while the small scale
structure constitutes a random realization of the assumed model (see
\citet{klypin_etal:lsc1} for a quantitative analysis).  The random
(unconstrained) waves dominate on small scales which leads to random
displacements of some large-scale features in the density distribution
by a few Mpc from realization to realization.  The realization used
for our simulations was selected from a set of low-resolution
dissipationless runs to have the position of the Virgo cluster to be
as close to its observed location as possible.  The realizations are
constructed assuming that the linear theory is valid on all scales at
the present epoch, and is then evolved backwards in time to the
initial epoch of the simulation by the linear growth factor.

A region within a radius of $30\mpch$ around the Virgo cluster inside
the $160\mpch$ computational box was selected for re-simulation at the
highest mass and force resolution.  The initial positions and
velocities for $1024^3$ particles were generated in the entire
$160\mpch$ box (roughly centered on the Virgo cluster).  Particles
outside the designated high-resolution region were then
merged\footnote{The larger mass (merged) particle is assigned a
velocity and displacement equal to the average velocity and
displacement of the smaller-mass particles.} into particles of larger
mass and this process was repeated for merged particles to construct a
nested hierarchy of particles of four different masses surrounding the
high-resolution region.  The simulation used four DM particle species
with DM particle mass of $3.16\times 10^{8}\Msunh$ in the highest
resolution region. The initial conditions used for our simulations are
identical to the simulation analyzed in \citet{klypin_etal:lsc1}.

The simulations presented here were run using $128^3$ zeroth-level
grid.  The initial gas density and velocity for each grid cell was
computed using DM particle positions and velocities and cloud-in-cell
(CIC) interpolation.  The subsequent mesh refinement was done using
criteria based on the local DM and gas densities. Namely, a cell of
level $L$ is marked for refinement if mass of DM particles (evaluated
using the CIC-smoothed density field) and/or mass of gas in the cell
exceeds specified threshold. The final refinement map is then created
by smoothing the original map using an algorithm similar to that
described in \citet{khokhlov98}. The smoothing generally enlarges
regions marked for refinement and reduces noisiness of the original
map.  The threshold can be chosen to be different at different levels
of refinement.  The density-based mesh refinement is appropriate as we
are interested in resolving collapsing high-density regions.  However,
the specific choice of threshold masses is arbitrary.

To ensure that the results of our study do not depend on the choice of
thresholds, we have run two simulations with very different numerical
criteria for refinement. In the first simulation (hereafter R$1$), the
mass thresholds were the same for all levels and were set to the
average mass of DM and gas per zeroth level cell ($\langle m_{\rm
DM}\rangle\approx 1.4\times 10^{11}\Msunh$ and $\langle m_{\rm
g}\rangle \approx 2.3\times 10^{10}\Msunh$, respectively). There was
no limit on the number of refinement levels; a total of 8 levels were
introduced in the highest density regions, corresponding to the
smallest cell size of $\approx 4.9\kpch$.  The second simulation
(hereafter R2) was started from the same initial conditions and used
the same threshold masses for refinement on the zeroth and the first
levels as the simulation R1. However, run R2 was limited to four
levels of refinement (corresponding to the smallest cell size of
$78\kpch$ and mass thresholds for levels 2 and 3 were set to $\langle
m\rangle/4$ and $\langle m\rangle/16$, respectively. The simulation
R2, therefore, has lower peak resolution than the R1. The fraction of
the volume resolved at the peak resolution is, however, much greater
in R2 than in R1. In particular, all regions of local overdensity
$\sim 20-30$ were refined to the peak resolution of $78\kpch$ in
R2. In the simulation R1 this resolution was reached only in regions
with local overdensity higher than $\sim 500$. For all the results
presented in this paper there is no significant difference between
simulations R1 and R2. In the remainder of the paper we will use
simulation R2 due to its higher resolution in the intermediate density
regions.

The time step on the zeroth level grid for the simulations,
$\Delta\tau_0$, was set at the beginning of each global step to
satisfy the Courant condition (with the Courant-Friedrichs-Lewy factor
of $cfl =0.4$) for cells on all levels and with the condition that the
corresponding step in expansion factor would not exceed $\Delta
a_0\leq 0.003$ (the latter condition is usually imposed only during
the early stages of evolution).  The time steps for particles and
cells on other levels were scaled appropriately: $\Delta\tau_l =
\Delta\tau_0 2^{-l}$. The simulations were started at $z=30$ and a
total of 364 global time steps were performed in both
simulations.

\section{Results}
\label{sec:results}

\subsection{Spatial distribution of gas and dark matter}
\label{sec:spdist}

At the current epoch most of the mass ($\approx 7.5\times
10^{14}\Msunh$) of the LSC is located in a filament roughly centered
on the Virgo cluster and extending over $\sim 40 \hmpc$.  The
simulated Local Group (LG) is located in an adjacent smaller filament,
which is not a part of the main body of the LSC, and has a peculiar
velocity of $\approx 250~\kms$ toward the Virgo cluster.  The peculiar
velocity flow in the vicinity of the LG in the simulation is
relatively ``cold'': the peculiar line-of-sight velocity dispersion
within $7 \hmpc$ of the LG is $\lesssim 60~\kms$, comparable to the
observed velocity dispersion of nearby galaxies. In the main filament
of the LSC, the peculiar velocities are higher with typical values of
$\sim 200-300~\kms$; the peculiar velocities are the highest, $\gtrsim
400-500~\kms$ in the immediate vicinity of the Virgo cluster.  The
overall matter distribution and peculiar velocity field in the the
simulated LSC region is discussed in detail in Paper I and we refer
reader to this paper for further information. Here we will briefly
discuss the distribution of gas and dark matter.

\begin{figure*}[tb!]
\epsscale{2.1}
\plotone{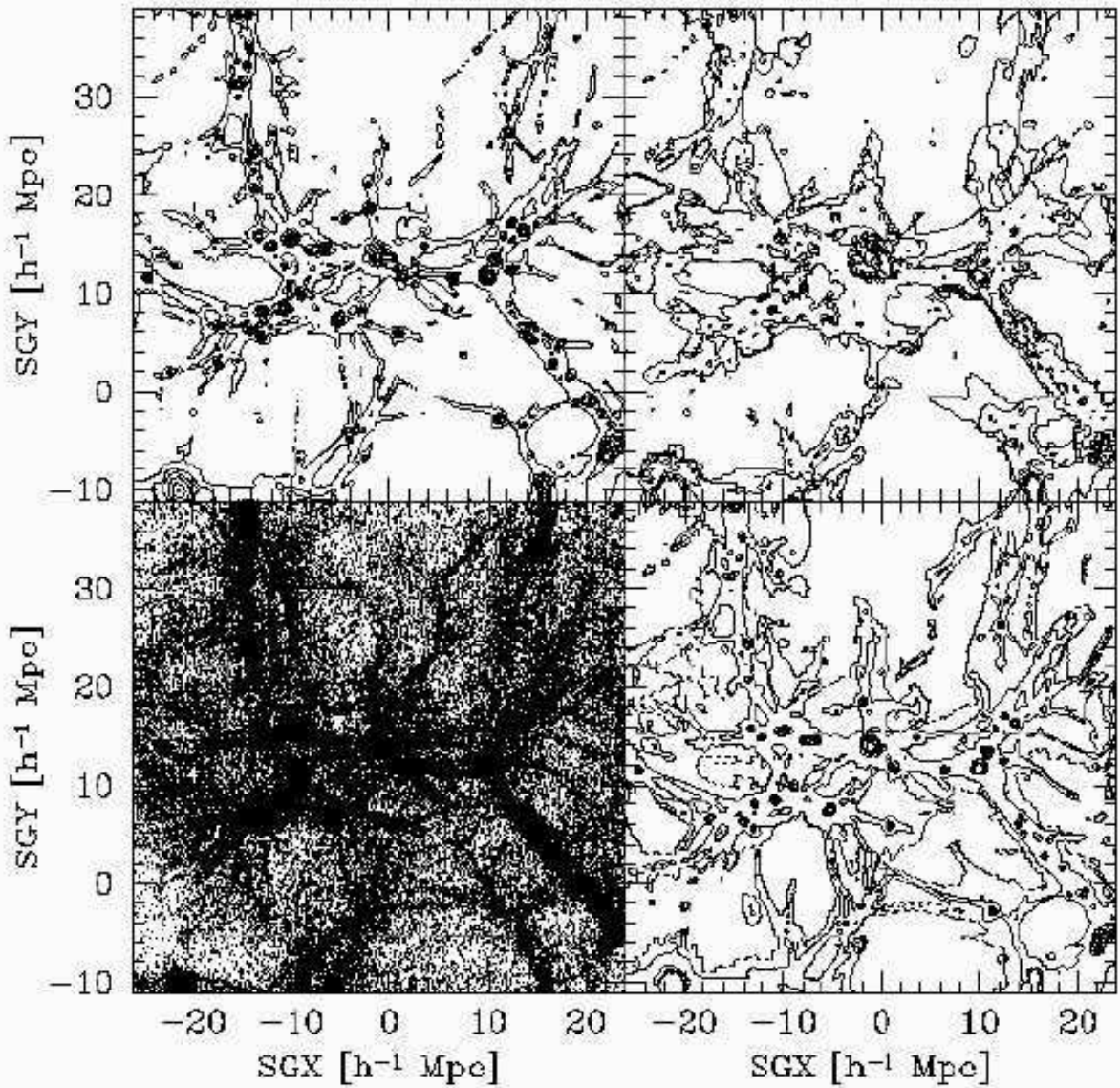}
\caption{\small
A slice through the gas and DM distribution in the simulation R2. The
slice is centered on the simulated Virgo cluster and is $15\hmpc$
thick. {\em Upper left panel:} the contour plot of the projected
density field, where the thick solid contours correspond to the mean
gas density, while subsequent contours correspond to the densities of
4, 8, 16, etc. times the mean density. {\em Lower left panel:} the
distribution of DM particles in the slice; randomly selected subset of
10\% of all the particles in the high-resolution region is shown. {\em
Upper right panel:} the contour plot of the emission-weighted
projected temperature field in the slice. The thick contour
corresponds to temperature of $3\times 10^5$~K, while other contours
correspond to temperatures of $10^6$~K, $4\times 10^6$~K, $6\times
10^6$~K, $8\times 10^6$~K, and $10^7$~K. {\em Lower right panel:} the
projected density fields of the Warm-Hot (solid contrours) and diffuse
phases of the gas (dotted contour). The thick solid contour
corresponds to the density of WHIM equal to the mean density of the
gas, while thin solid contours correspond to the WHIM densities of 8,
16, 32, etc. times the mean gas density; the dotted contour
corresponds to the density of diffuse gas equal to $0.1$ of the mean
gas density.
}\label{fig:slice4}
\end{figure*}

Figure~\ref{fig:slice4} shows a $15\hmpc$ slice through the gas and DM
distribution in the simulation R2. The slice is centered on the
simulated Virgo cluster. The left panels show the (continuous) density
field of gas (upper left panel) and distribution of DM particles
(lower left panel). Comparison of the gas and DM distributions shows
that at these large scales gas traces the gravitationally dominant DM
distribution very well: similar filamentary structures and halos can
be identified in both panels. As discussed above, the dominant
structures are the ``Virgo'' cluster in the center and the horizontal
filament around it. The right panels show the projected
emission-weighted temperature field of the gas (upper panel) and
projected density of the Warm-Hot (solid contours) and diffuse (dotted
contour) phases of the gas.  The panels show that warm ($\sim
10^5-10^6$~K) gas is located in filaments, while hotter gas ($>5\times
10^6$~K) is concentrated around virialized halos of groups and
clusters. It is the Warm-Hot gas that will be the main focus of our
study.

Below we discuss properties and possible observational signatures of
the intergalactic gas in the constrained simulations of the LSC
region.  The analysis presented below excludes the inner structure of
the Virgo cluster.  The cooling and feedback processes neglected in
our simulations would affects the structure of the cluster. We will
present a constrained simulation of the Virgo cluster that includes
heating and cooling processes in a forthcoming paper.

\begin{figure}[tb!]

\epsscale{1.2}
\plotone{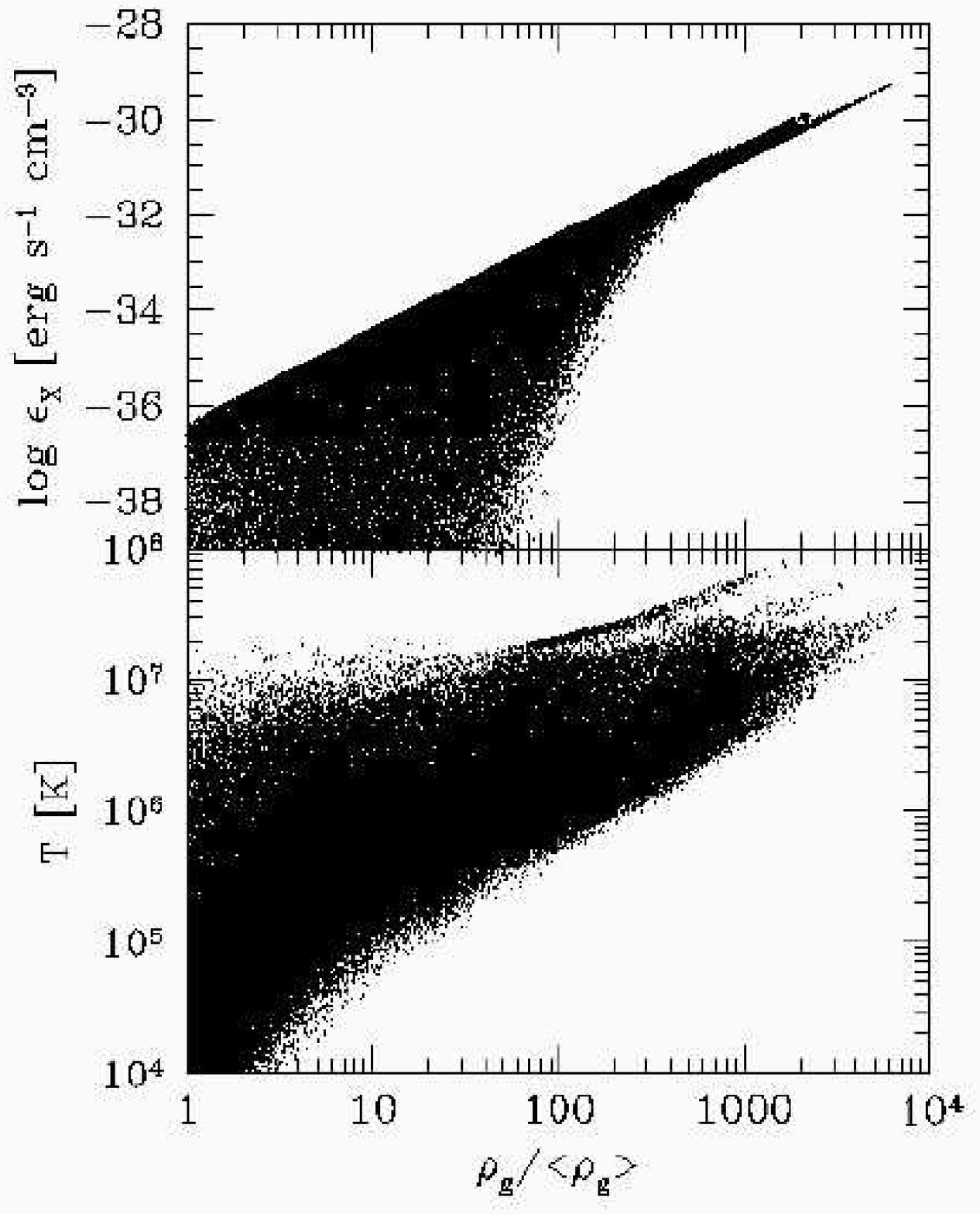}
\caption{\small X-ray (0.5-2 keV) emissivity ({\em top panel}) and gas temperature ({\em bottom
  panel}) as a function of gas density. The gas density is given in
  units of the mean density. The X-ray emissivity was computed using
  the \citet{raymond_smith77} code assuming uniform metallicity of
  $0.3$ solar.  Points correspond to randomly selected individual mesh
  cells in the high-resolution simulation region. Note that the cells
  plotted correspond to relatively high-density regions and represent
  less than half of the volume; the median values of gas density,
  temperature, and X-ray emissivity by volume (i.e., half of the
  volume is at values below the median) are 0.29, 5000~K (the minimum
  assumed gas temperature), and $10^{-42}{\ \rm ergs\ s^{-1} cm^{-3}}$.
  }\label{fig:dte}
\end{figure}

\subsection{Properties of gas}

\citet{dave_etal01} present a detailed discussion of gas properties
in the currently favored structure formation models. They classify the
gas in four phases: diffuse ($\delta <1000$, $T<10^5$ K), condensed
($\delta >1000$, $T<10^5$ K), hot ($T>10^7$~K), and warm-hot
($10^5<T<10^7$~K), where $\delta$ and $T$ are gas overdensity and
temperature, respectively.  The simulations presented here do not
include cooling and thus cannot predict the fraction of gas in
condensed phase, while fraction of gas in the hot phase is sensitive
to the presence or absence of a small number of galaxy clusters in the
simulation volume. However, the hot gas fraction is expected to be
small ($\sim 5\%$) and the gas not in diffuse or WH phase should
correspond approximately to the condensed fraction in the simulations
with cooling.  It is therefore interesting to compare the fraction of
WH and diffuse gas to those in previous simulations.
\citet{dave_etal01} find that $\sim 40\%$ and $\sim 30\%$ of
the gas at present epoch is in the diffuse and Warm-Hot (WH) phases,
respectively.  The fraction of WH gas appears to be rather robust:
approximately the same fraction of WH gas is measured in simulations
with and without cooling and/or supernova feedback.  The corresponding
fractions of diffuse and WH gas in the high-resolution region of our
simulations are $\Omega_{\rm diff}/\Omega_b\approx 0.41$ and
$\Omega_{\rm WH}/\Omega_b\approx 0.34$, respectively.  Both fractions
are in good agreement with the fractions found by
\citet{dave_etal01}.

Figure~\ref{fig:dte} shows the X-ray volume emissivity in the
$[0.5-2]$~keV energy range and gas temperature as a function of gas
density in the simulation R1. The points in the figure represent
individual mesh cells randomly selected from the high-resolution
region of the simulation (see \S~\ref{sec:sims}). The X-ray emissivity
was calculated using the \citet{raymond_smith77} code assuming
complete ionization and uniform gas metallicity of $0.3$ solar.
Figure~\ref{fig:dte} shows that temperature is correlated with gas
density, although there is much scatter (especially at low gas
densities). As noted by \citet{dave_etal01}, the average can be
roughly described as $\rho\propto T$.  The large scatter in
temperature translates into significant scatter in X-ray emissivities.
The relatively sharp upper edge in the $\epsilon_X-\rho_g$ scatter
plot is due to the fact that at high tmperatures (corresponding to the
highest emissivities), the emissivity is a strong function of gas
density ($\propto \rho_g^2$) and is a relatively weak function of
temperature ($\propto T^{1/2}$). As can be seen from the lower panel,
the maximum temperatures of the gas change by less than a factor of
ten over four orders of magnitude in density. The upper envelope in
the upper panel is therefore dominated by the density dependence:
$\epsilon_x \propto \rho_g^2$ and scatter in temperature intoduces
only a small scatter at the maximum emissivity at a given $\rho_g$.

Note that although a large fraction of WH gas is at relatively low
overdensities ($\rho_g/\langle\rho_g\rangle\lesssim 100$), the X-ray
emission is dominated by gas at higher overdensities. Note also that
at overdensities of $\rho_g/\langle\rho_g\rangle\sim 10$ most of the
gas has temperatures in the range $T\sim 10^5-10^6$~K.  The
correlation of emissivity and temperature with density presented in
Figure~\ref{fig:dte} are in good agreement with results of previous
studies \citep[][see their Fig.~6 and Fig.~2,
respectively]{dave_etal01,croft_etal01}.

\subsection{Sky maps}
\label{sec:maps}

The use of observational constraints implies that the computational
box axis in our simulations are mapped with the same supergalactic
coordinates as the observed universe.  Because observational
constraints small scales were not too tight, there was a room for
fine-tuning the position of the origin of coordinates (i.e., position
of the Local Group in the simulations).  We choose the origin to be
the location of a pair of galaxy size halos in the high-resolution
dissipationless simulation described in Paper I (see Fig.~6 in that
paper). For this choice of origin, the simulated Virgo cluster has
supergalactic coordinates of $(SGX, SGY, SGZ)=-1.2,13.7,0.7\Mpch$ and
is located $13.74\Mpch$ from the Local Group.  Knowledge of the
supergalactic coordinates allows us to construct sky maps in real
celestial coordinates.

\begin{figure}[tp]
\begin{center}
\includegraphics[angle=-90,scale=0.4]{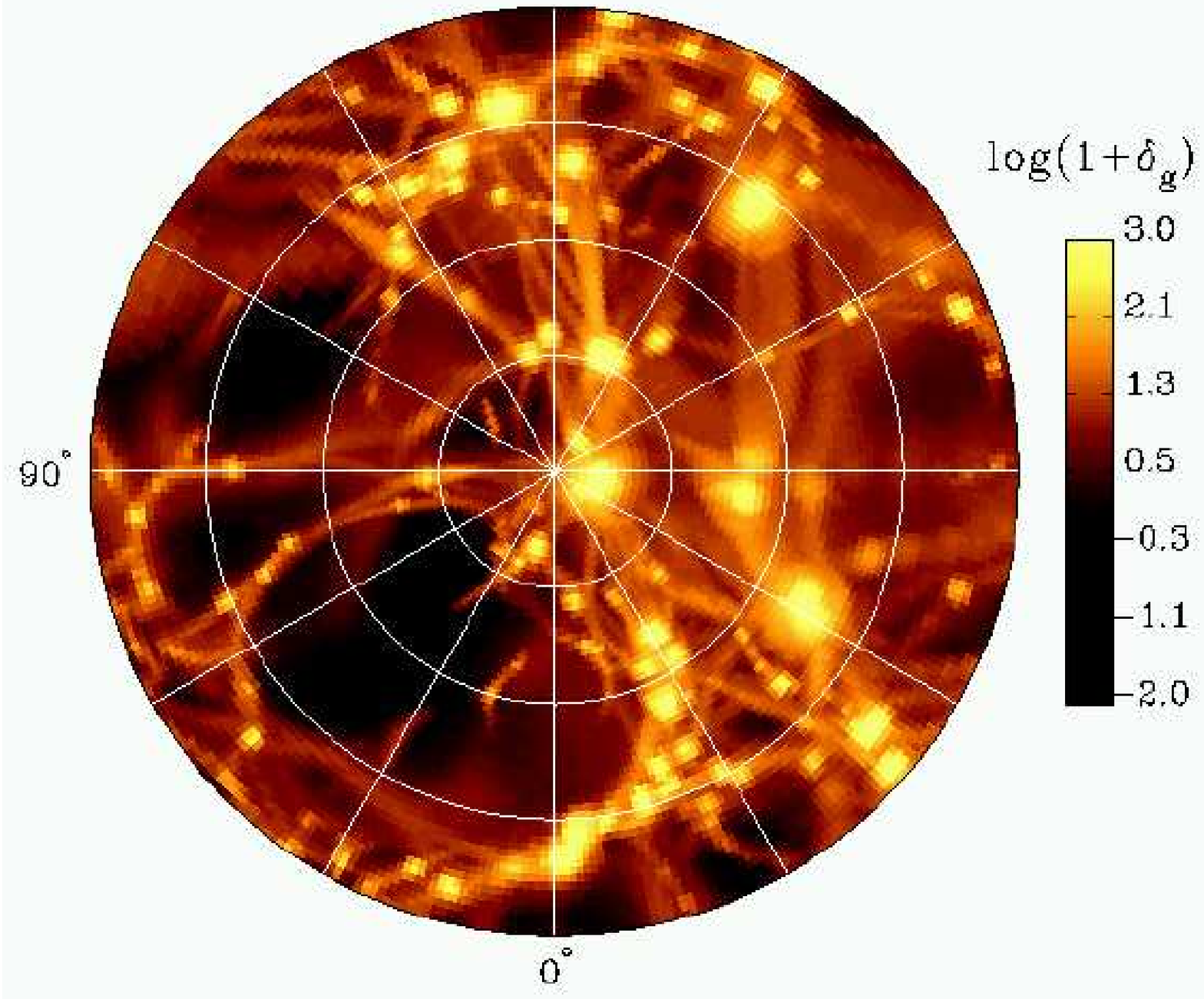}
\includegraphics[angle=-90,scale=0.4]{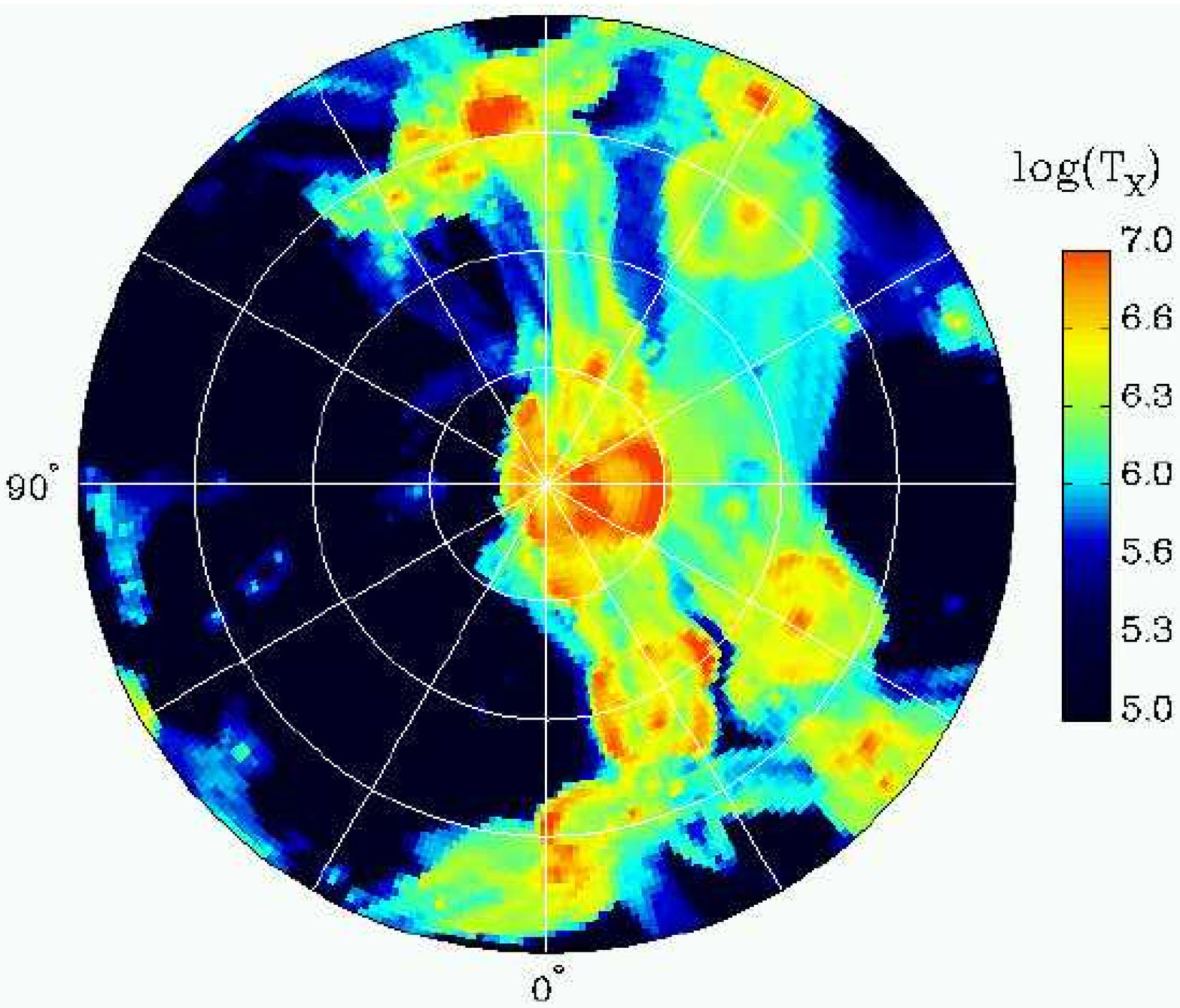}
\end{center}
\caption{\small Sky maps of density ({\em top panel}) and X-ray
emission weighted temperature ({\em bottom panel}) of the gas in the
simulated Local Supercluster.  The panels show polar views around the
North Galactic Pole ($b>30^{\circ}$).  Radial lines and circles
correspond to spacings of $30^{\circ}$ and $15^{\circ}$ in galactic
$l$ and $b$, respectively; $l=0^{\circ}$ is at the bottom and
$l=90^{\circ}$ is on the left; $b=90^{\circ}$ is in the center.  The
density represents the average maximum overdensity along a line of
sight in a given map ``pixel''; the temperature is in degrees Kelvin
and is avaluated as an emission-weighted average along lines of sight
in a ``pixel'' spanning from $2\hmpc$ to $25\hmpc$ from the origin.
The corresponding color scales are shown by the bar to the right of
each panel.  The Virgo Cluster is close to the center of the plots;
the Ursa Major cluster is close to the top of the plots.  The Local
Void is in the bottom left quadrant.  }\label{fig:sky1}
\end{figure}

In the following we present a series of sky maps showing projections
of various quantities in a region of the sky around the north galactic
pole. The maps are shown as polar views in the galactic coordinates,
with the North Galactic Pole (NGP; $b=90^{\circ}$) in the center. We
plot only a region of $b>30^{\circ}$. We do not consider other regions
of sky because 1) most of the nearby structures in the high-resolution
region ($\approx 30\mpch$ around the Virgo cluster) of the simulations
are concentrated around the NGP (the Virgo cluster is located within
$15^{\circ}$ of the NGP) 2) the sky at low galactic latitudes is more
difficult to observe due to galactic obscuration.  All maps include
only cells in the high-resolution region of the simulation; the
resolution outside this region was considerably worse (cell size of
$1.25\mpch$) and our predictions would be considerably less accurate.
All quantities represent averages over a number of random lines of
sight passing through each map pixel. The integration along lines of
sight was done from $2\hmpc$ to $25\hmpc$. We thus exclude the regions
in the immediate vicinity of the LG which are not modelled accurately
in our simulations.

Figure~\ref{fig:sky1} shows the sky maps of the gas overdensity
and X-ray emission weighted temperature. Note that although the gas
distribution is very filamentary in appearance, the overall
distribution is flattened along a plane which is roughly aligned with
the observed supergalactic plane. The large void in the lower left
corner is the counterpart of the Local Void in the observed
distribution of nearby galaxies.  The temperature of gas in filaments
ranges from one to a few million degrees K and reaches tens of million
K within and around the virialized regions of groups and
clusters. Note that these regions are surrounded by the narrow regions
of enhanced temperature corresponding to accretion shocks. The
accretion shocks are also visible around massive filaments. The two
hottest regions in the center and at the top of the plot are the
environments of the Virgo and Ursa Major clusters, respectively.

\begin{figure*}[tp]
\includegraphics[angle=-90,scale=0.4]{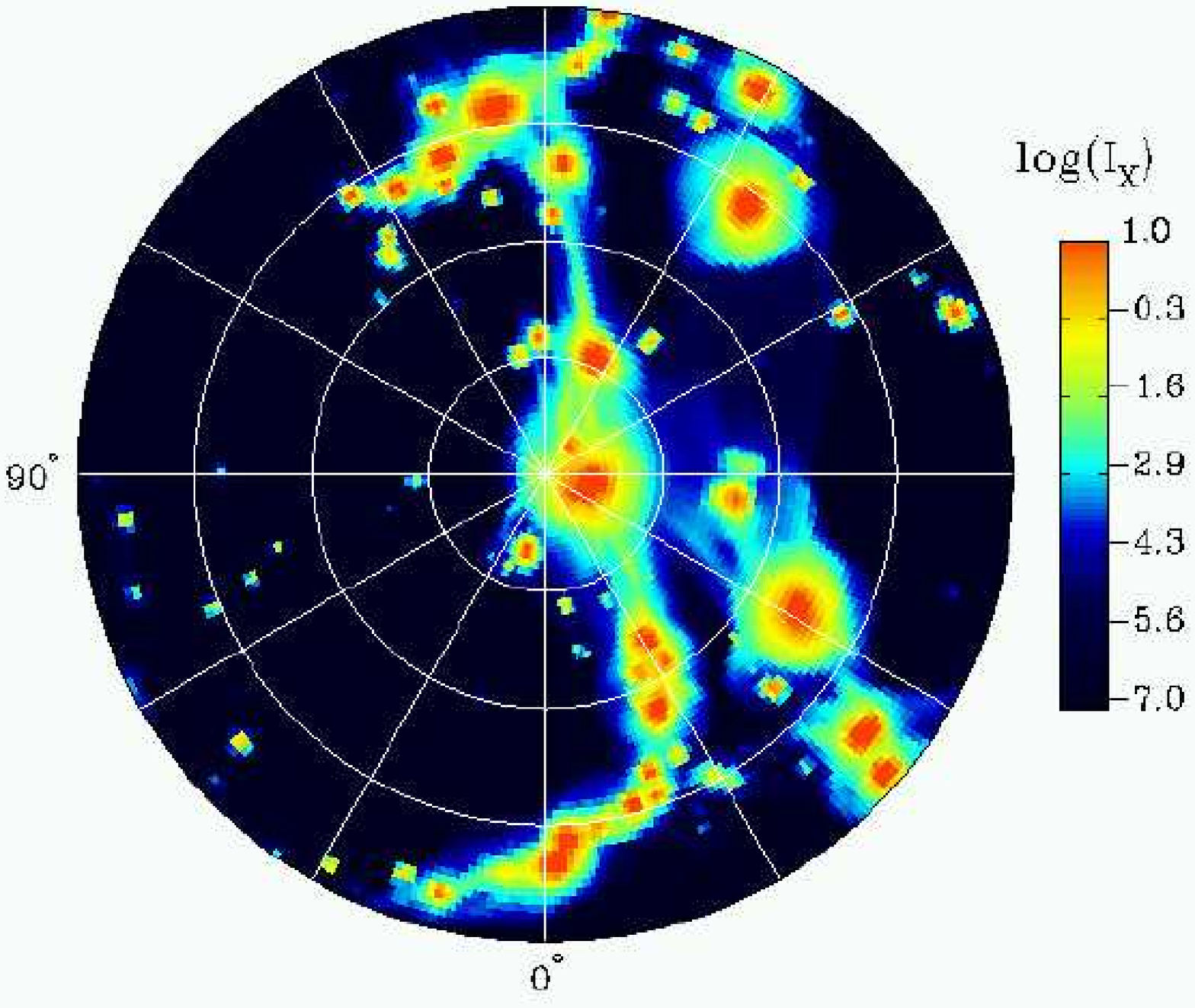}
\includegraphics[angle=-90,scale=0.4]{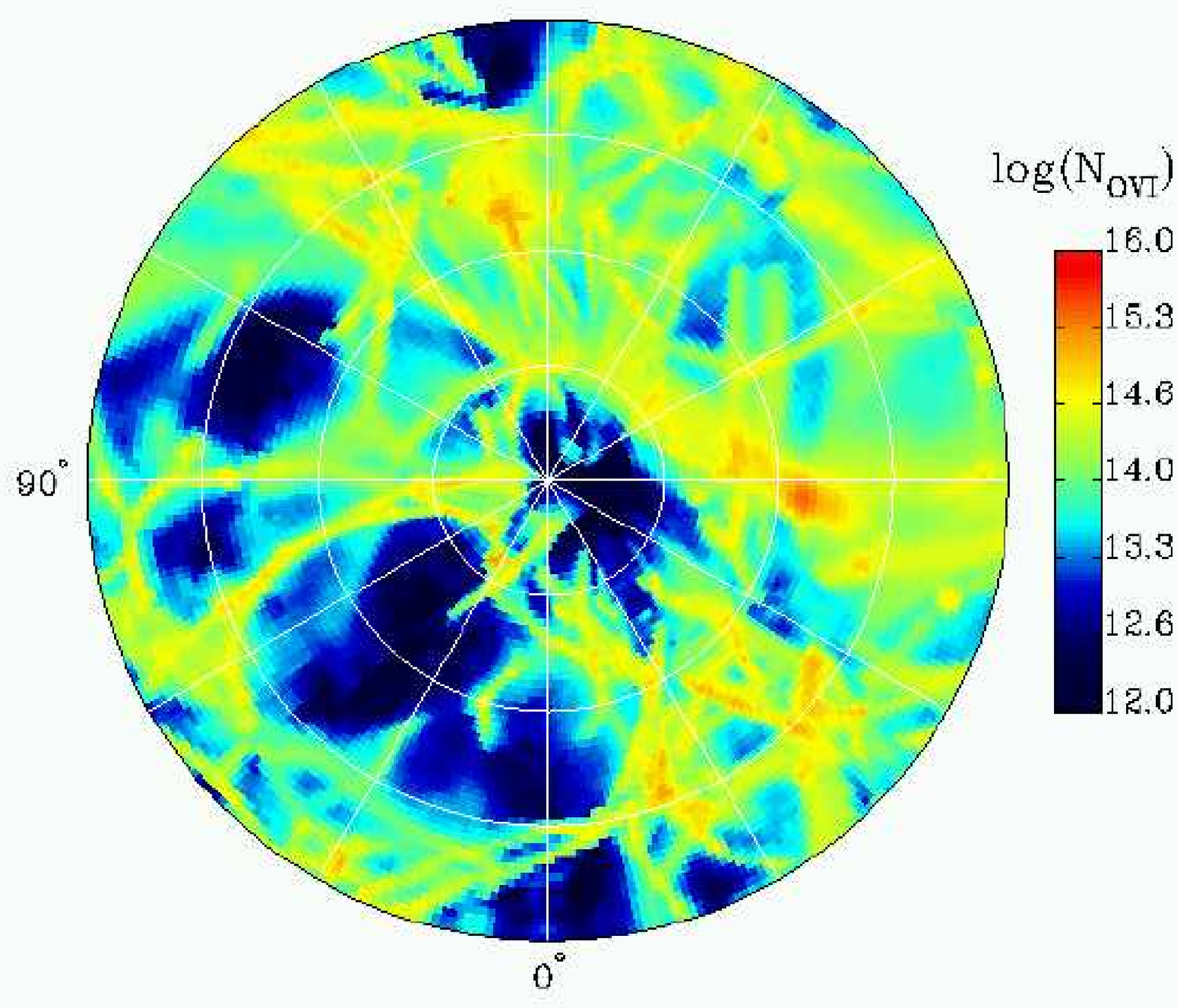}
\includegraphics[angle=-90,scale=0.4]{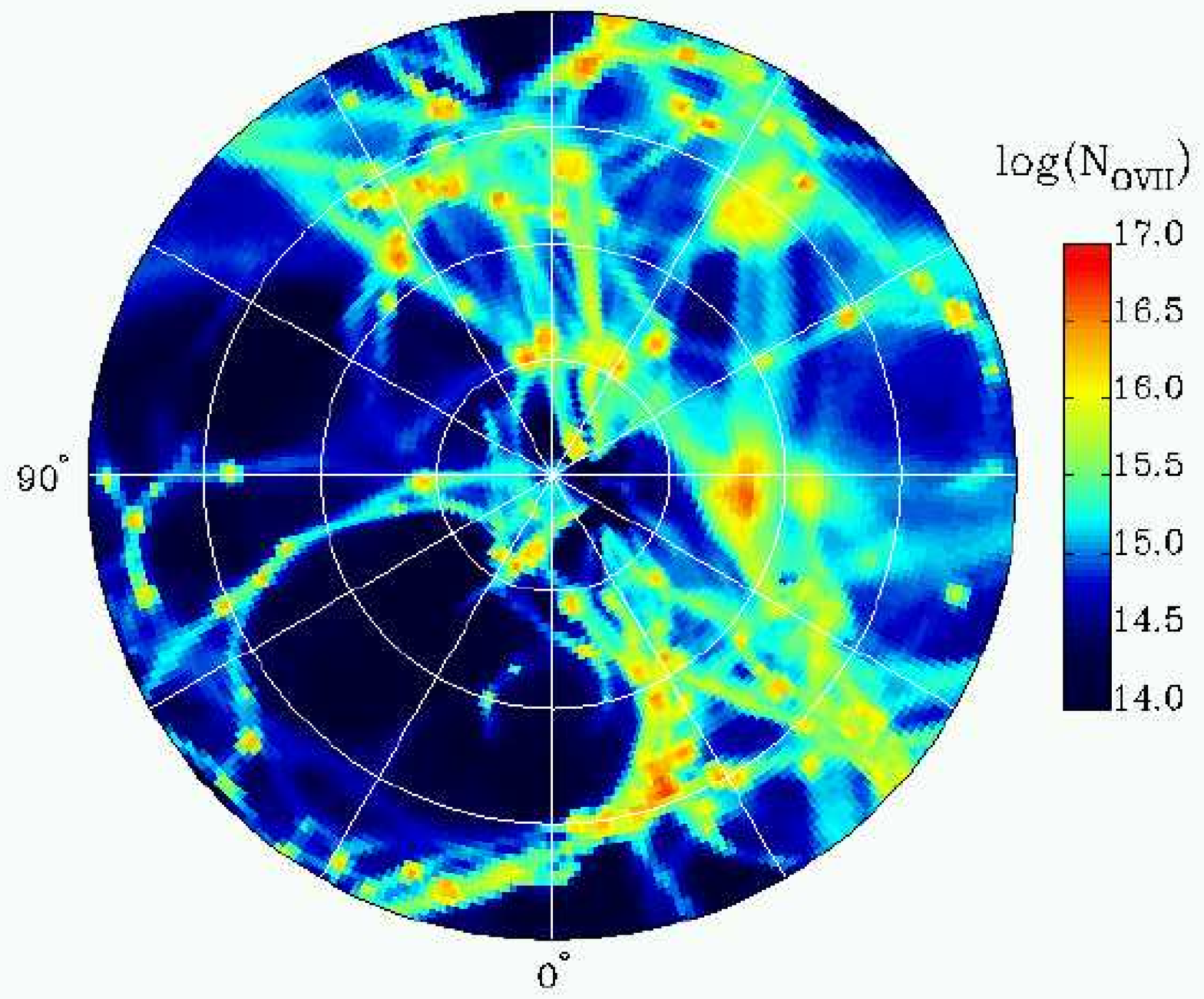}
\includegraphics[angle=-90,scale=0.4]{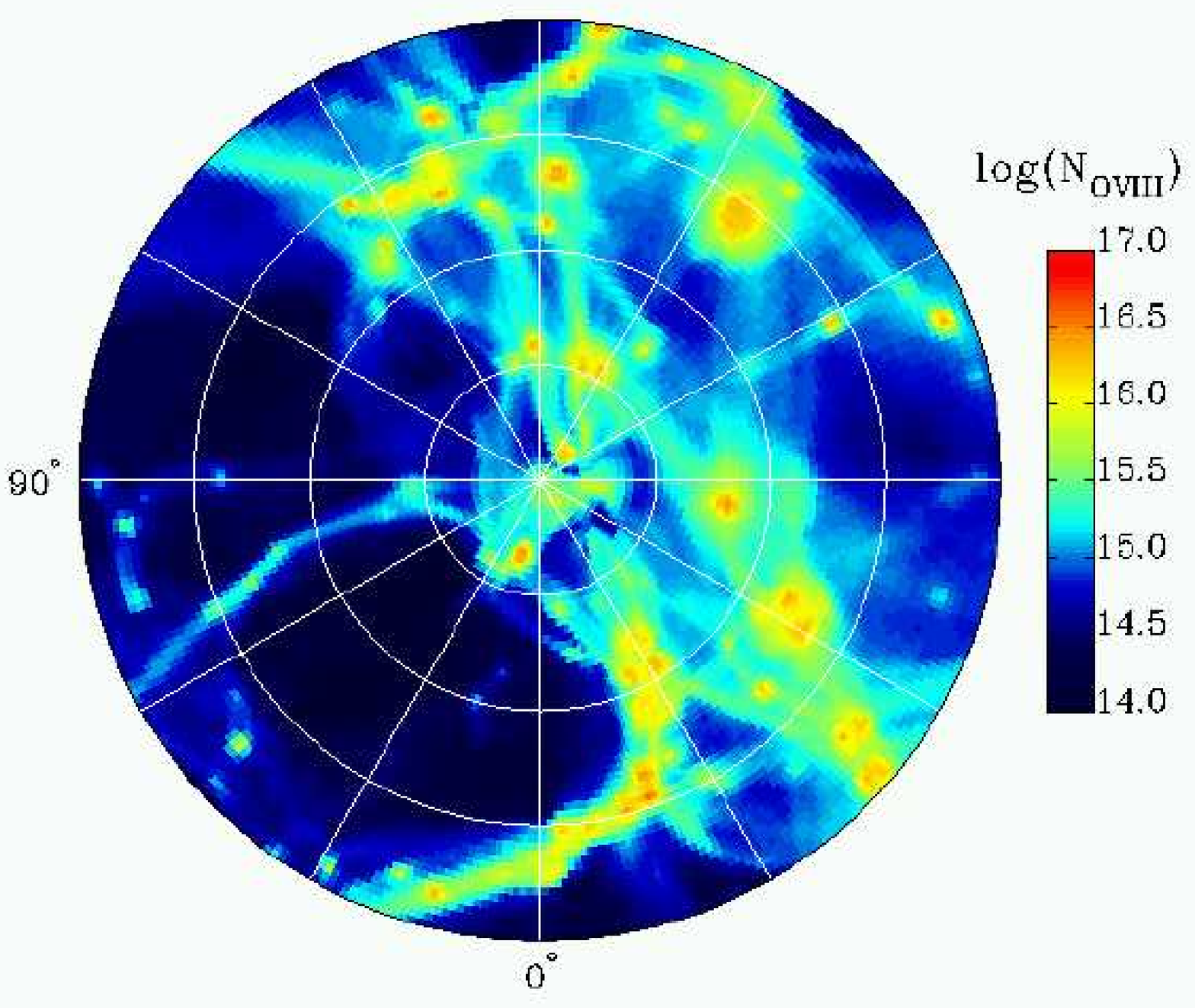}
\caption{\small Sky maps of X-ray surface brightness in $[0.5-2]$~keV 
({\em top left}) and column densities of oxygen ions: OVI ({\em
top right}), OVII ({\em bottom left}), and OVIII ({\em bottom right}).
The X-ray brightness is given in ${\rm keV s^{-1} cm^{-2} sr^{-1}
keV^{-1}}$, while column densities are in ${\rm cm^{-2}}$.  The layout
is similar to that of Figure~\ref{fig:sky1}. The brightness and column densities
were evaluated by averaging along random lines of sight within each map ``pixel''
in the distance range $[2,25]\hmpc$. See text for details.
}\label{fig:sky2}
\end{figure*}

Figure~\ref{fig:sky2} shows sky maps of surface brightness of X-ray
emission in $[0.5-2]$~ keV energy band and column densities of oxygen
ions: OVI, OVII, and OVIII. The hot, $T\gtrsim 10^6$~K gas in
filaments and virialized regions (see Figures~\ref{fig:dte} and
~\ref{fig:sky1}) should emit soft X-ray radiation. As can be seen from
the Figure~\ref{fig:sky2}, the X-ray surface brightness map has a more
patchy appearance than both the density and temperature maps.  This is
because X-ray emission is roughly proportional to $\rho^2T^{0.5}$ and
is thus heavily weighted towards the highest density regions.

\begin{figure}[t]
\includegraphics[scale=0.4]{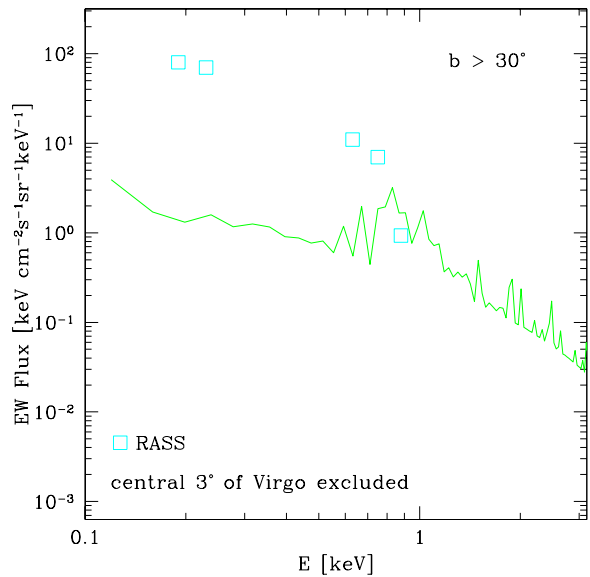}
\caption{\small The X-ray spectrum of gas in the LSC region. The
spectrum, shown by the green solid line, represents the
emission-weighted average over the sky region around the NGP
($b>30^{\circ}$) shown in Figure~\ref{fig:sky2}. Open squares show the
broad-band measurements of the intensity of the X-ray background by
the RASS with known contributions of AGNs, stars, and the Local Hot
Bubble subtracted \citep{kuntz_etal01}. The $3^{\circ}$ region around
the center of the Virgo cluster was excluded in both model and
observations.  }\label{fig:xspec}
\end{figure}

The emission-weighted average X-ray spectrum from the considered sky
region is shown in Figure~\ref{fig:xspec}. The spectrum was computed
using the \citet{raymond_smith77} plasma code and assuming uniform gas
metallicity of 0.3 solar. The model spectrum is compared to the
broad-band measurements of soft X-ray extragalactic background from
the RASS survey with the contribution of AGNs, stars, and the Local
Hot Bubble subtracted \cite[see][for details]{kuntz_etal01}.  In both
models and observations emission from the $3^{\circ}$ region around
the Virgo cluster was excluded.  The figure shows that energies
$\lesssim 0.5$~keV the predicted emission of the intergalactic gas is
significantly lower than the measured background.  As argued by
\citet{kuntz_etal01}, this emission is likely due to the Milky Way's
own hot halo. However, at energies $\sim 1$~keV the X-ray emission of
the intergalactic gas, although still relatively small ($\sim 10\%$ of
observed background at these energies; the measured total background
at $1$~keV is $\sim 10\ {\rm keV\ s^{-1} cm^{-2} sr^{-1} keV^{-1}}$),
may be important for understanding the total background at these
energies.

In addition to the X-ray brightness map, Figure~\ref{fig:sky2} shows
the sky maps of column densities of three ionic species of oxygen:
OVI, OVII, and OVIII. The concentrations of these species were
calculated using the current version (94.00) of the CLOUDY code
\citep{ferland_cloudy94} assuming that the gas is embedded in the
observed XRB spectrum and has uniform metallicity of 0.3 solar.  We
assume that power-law XRB spectrum.  $P(E)=AE^{-\Gamma}$, with $A=9.5\
{\rm keV\ s^{-1} cm^{-2} sr^{-1} keV^{-1}}$ \citep{kuntz_etal01} and
$\Gamma=1.46$ \citep{chen_etal97}; this is equivalent to the flux
$J_{\nu}=1.2\times 10^{-26}(\nu/\nu_X)^{-0.46}\ {\rm
ergs\ s^{-1}cm^{-2}sr^{-1}Hz^{-1}}$, where $h\nu_X=40\ {\rm keV}$.  We
also assume that the spectrum has an exponentials cutoff at energies
above $h\nu_X$ and at very small energies ($\ll 1$~keV).

Comparison of the column density maps in Figure~\ref{fig:sky2} to the
projected density map in Figure~\ref{fig:sky1} shows that high column
density regions are roughly correlated with regions of high gas
density.  Note, in particular, that high-column density regions are in
general located near the supergalactic plane.  Nevertheless, different
ionic species trace regions of somewhat different densities and
temperatures \citep[e.g.][]{hellsten_etal98,nahar99} and their maps,
therefore, differ in details. For instance, the column densities of
OVI and OVII are depressed in the vicinity of the Virgo cluster
($l\sim 300^{\circ}$, $b\sim 85-90^{\circ}$) due to the high gas
temperatures around the Virgo cluster. The decrease in column density
of OVIII in this region is relatively small.

Our primary interest is evaluating the prospects of detection of the
gas in the LSC region. The oxygen is the most abundant of the metals
that can absorb or emit in UV and X-rays. The absorption or emission
by lines of the ionic species of oxygen is therefore the most
promising probe of the LSC gas. The strongest lines of OVII and OVIII
are in soft X-rays at $0.5740$~keV and $0.6536$~keV, respectively. The
OVI can be detected using resonance line doublet in UV at
$\lambda\lambda 1031.92,\ 1037.62\ {\rm\AA}$.  The current instruments
in X-ray ({\sl Chandra and XMM-Newton}) and UV ({\sl FUSE}) are
sufficiently sensitive to detect at least some of the intergalactic
gas. The exact detection limit depends on the brightness of the
background source (in the case of absorption) and specifics of the
instrument and observation (time of integration, spectral resolution,
etc.). For an X-ray bright quasar and a reasonable integration time
(100 ks), the \citet{fang_canizares00} estimate that {\sl Chandra},
{\sl XMM}, and the planned {\sl Constellation-X} should detect OVIII
(sensitivity to OVII is similar) at column densities higher than
$3.5\times 10^{16}$, $7.6\times 10^{15}$, and $1.3\times 10^{15}\ {\rm
cm^{-2}}$, respectively. The OVI was robustly detected in absorption
with {\sl FUSE\/} for column densities $\gtrsim 10^{14}\ {\rm
cm^{-2}}$ \citep{tripp_etal00}.

The column density maps in Figure~\ref{fig:sky2} show that for the
assumed metallicity the column densities of OVI should be detectable
with {\sl FUSE} over a large fraction of the sky, especially near the
supergalactic plane. On the other hand, the column densities of OVII
and OVIII detectable with {\sl Chandra} and {\sl XMM-Newton} cover a
much smaller fraction of the sky. This fraction is much higher for the
sensitivity of the {\sl Constellation-X}.

\begin{figure}[t]
\epsscale{1.0}
\plotone{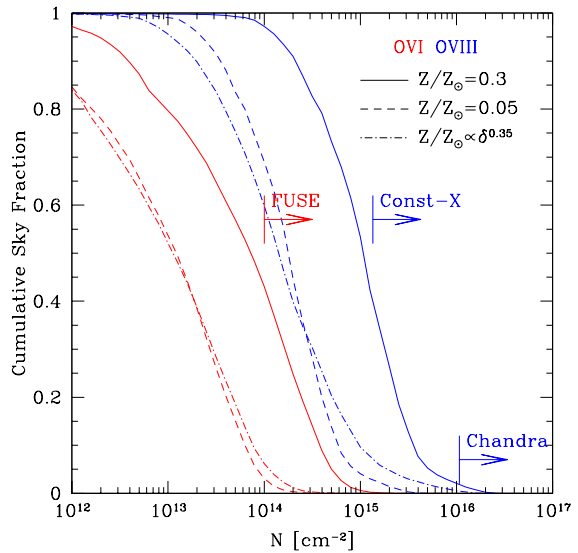}
\caption{\small The fraction of the sky around the NGP
($b>30^{\circ}$) at column densities greater than a given value $N$
for OVI and OVIII for different metallicity models.  The results for
OVII are quantitatively similar to those for OVIII. The {\em solid\/}
and {\em dashed\/} lines show results for the uniform metallicity of
gas of 0.3 and 0.05 solar, respectively; the {\em dot-dashed\/} shows
results for the density-dependent metallicity of the gas:
$Z/Z_{\odot}=0.025\delta_g^{0.35}$, where $\delta_g$ is gas
overdensity, which approximates results of numerical simulations of
\citet{co99_chemical}. The sensitivity of current and future
instruments is also shown (sensitivity of the XMM-Newton is
intermediate to those of Constellation-X and Chandra).  Note that for
the metallicity of the IGM of 0.3 solar, OVI and OVII/OVIII have
column densities detectable with the FUSE and Constellation-X over a
large fraction of this sky region.  }\label{fig:skyfrac}
\end{figure}

The cumulative fraction of the sky (estimated over the region of the
sky shown in Figure~\ref{fig:sky2}, i.e. at $b>30^{\circ}$) at column
densities above a given value is shown in Figure~\ref{fig:skyfrac}.
This fraction can be thought of as a probability to have a detectable
column density in a random line of sight. The figure shows predictions
under three different models for gas metallicity: uniform metallicity
of 0.05 and 0.3 solar, and density-dependent law,
\begin{equation}
Z/Z_{\odot}=0.025(\rho_g/\langle\rho_g\rangle)^{0.35},
\label{eq:co}
\end{equation}
which approximates results of numerical simulations of
\citet{co99_chemical}. For uniform metallicity of $Z/Z_{\odot}=0.3$,
the probability is $\approx 0.5$ for the detection of OVI absorption
with {\sl FUSE}.  The OVIII absorption (results for OVII are
quantitatively similar) should be detected with the probability of
$\approx 0.5$ with {\sl Constellation-X} and $\lesssim 0.05$ with the
{\sl XMM-Newton} and {\sl Chandra}. For lower metallicities of the
gas, the probabilities are considerably lower. Note that sky fraction
distributions for the density-dependent metallicity are similar to the
uniform metallicity of $Z/Z_{\odot}=0.05$. This indicates that the
signal is dominated by gas at overdensities of $\sim 5-10$ (see
eq.~\ref{eq:co}) and is sensitive to the average metallicity of this
gas.  Note that the fractions are calculated over the entire sky
region around the NGP ($b>30^{\circ}$), which includes sizeable voids
in the gas density; the fractions will be higher if one focuses on the
sky regions near the supergalactic plane.

\section{Discussion  and conclusions}
\label{sec:discussion}

We presented results of high-resolution gas dynamics$+N$-body
cosmological simulations of the Local Supercluster region constrained
to reproduce the observed nearby large-scale density field. The
simulations focused on the region of $\approx 30\mpch$ around the
Virgo cluster and followed the dynamics of dissipationless dark matter
and adiabatic dynamics of gas using the new version of the Adaptive
Refinement Tree code that includes gasdynamics. The detailed
discussion of constrained simulations, setup of initial conditions,
and properties of the density and velocity fields in the simulated
Local Supercluster are presented in a separate paper
\citep{klypin_etal:lsc1}. In this paper we use the simulations to
study the properties and possible observational signatures of
intergalactic medium in the LSC region.  We find that within $30\mpch$
of the Virgo cluster $\approx 30\%$ of baryons is in the warm/hot
phase at $T\sim 10^5-10^7$~K and $\approx 40\%$ are in the diffuse
phase ($T<10^5$~K, $\delta<1000$). The latter phase occupies most
($\gtrsim 90\%$) of the volume.

We presented simulated sky maps of $0.5-2$~keV X-ray surface
brightness and column densities of three ionic species of oxygen (OVI,
OVII, and OVIII).  The maps are constructed in real galactic
coordinates for the sky region around the North Galactic Pole
($b>30^{\circ}$). The map of X-ray brightness has a patchy appearance
with most of the emission coming from the high-density environments of
groups and clusters. We compared emission-weighted average X-ray
spectrum from the warm/hot gas in the simulation
(Figure~\ref{fig:xspec}) to the measurements of soft X-ray background
\citep{kuntz_etal01}.  The predicted flux at energies $\lesssim
0.5$~keV is $10$ to $100$ times lower than the observed flux. The LSC
gas therefore cannot explain this soft X-ray emission which is likely
to be associated with the hot halo of the Milky Way. At energies
around $1$~keV, the predicted flux constitutes $\approx 5-10\%$ of the
total XRB flux and is therefore an important component of the XRB at
these energies. At higher energies, the contribution of the LSC gas to
the XRB is insignificant.  It is not clear whether the X-ray emission
of the intergalactic gas can be reliably detected in the near future.
Such detection is difficult because one needs a survey that covers a
large sky area and is sufficiently sensitive to detect the signal only
$\sim 1\%$ of the XRB. The diffuse LSC emission should be correlated
with the supergalactic plane, but so is the X-ray emission from nearby
AGNs \citep{shaver_pierre89}.

Recently, \citet{boughn99} analyzed the {\sl HEAO 1} full sky
$2-10$~keV X-ray map and found evidence for the ``diffuse''
(unresolved) X-ray emission associated with the supergalactic plane.
The quoted maximum surface brightness of the diffuse emission
constitutes about $1\%$ of the $2-10$~keV XRB and is $I_{\rm X}\approx
5\times 10^{-10}\ {\rm ergs\ s^{-1} cm^{-2} sr^{-1}}$ (see his
Table~1). This corresponds to $\approx 0.04\ {\rm keV}$ ${\rm s^{-1}}$
${\rm cm^{-2} sr^{-1} keV^{-1}}$ assuming that flux is constant over
$2-10$~keV.  This flux is consistent with the predicted flux from the
WH gas in our simulation (see Fig.~\ref{fig:xspec}): $10^{-2}-10^{-1}
{\rm keV\ s^{-1} cm^{-2} sr^{-1} keV^{-1}}$ at energies $2-5$~keV
(i.e., $\approx 0.1-1\%$ of the total XRB at these energies; see
\citealt{kuntz_etal01}); the flux decreases steeply at higher
energies.

As can be seen from the X-ray brightness map in Figure~\ref{fig:sky2},
most of the X-ray emission is indeed concentrated towards the
supergalactic plane. However, the emission is patchy and is far from
being uniform. In contrast, \citet{boughn99} modelled the gas
distribution using a simple ``pillbox'' model for the distribution of
gas in the LSC region: the uniform gas distribution within a disk of
radius $R_{\rm SC}$ and thickness $H_{\rm SC}\approx 0.25R_{\rm SC}$.
For the detected diffuse X-ray flux this model implies gas density of
$2.5\times 10^{-6}{\ \rm cm^{-3}}$ for temperatures of around
$10$~keV.  This temperature is much higher than the typical
temperatures of the LSC gas in our simulation. The difference is due
to the fact that the gas distribution in the simulated LSC is not
described by the pillbox model.  As can be seen from
Figures~\ref{fig:sky1} and \ref{fig:sky2}, the distribution of matter
in the LSC region is filamentary rather than disk-like and the X-ray
emission is far from being uniform. The bulk of the X-ray emission
thus comes from the relatively high-density, $n_e\sim
10^{-5}-10^{-3}{\ \rm cm^{-3}}$, regions within and around groups and
clusters.

The hot and dense regions of the LSC should also distort the cosmic
microwave background radiation via the inverse Compton or Doppler
scattering, the thermal and kinetic Sunyaev-Zel'dovich effect,
respectively \citep[SZ;][and references therein]{SZ80}. The
temperature fluctuations of the CMB due to the non-relativistic
thermal SZ effect can be written as:
\begin{equation}
\frac{\Delta
T}{T}=y\left(\frac{e^x+1}{e^x-1}-4\right),
\end{equation}
where $x\equiv hv/kT_{\rm CMB}$, $y=1.117\times 10^{-34}\int n_eT_edl$
(all quantities are in cgs units and the integral is along
line-of-sight). In the Rayleigh-Jeans regime ($hv\ll kT_{\rm CMB}$):
$\Delta T/T\approx 2y$; the deviation of CMB temperature along a given
direction is thus proportional to the gas pressure integrated along
this direction.

Although for temperatures and densities typical for the gas in the LSC
the SZ effect is not expected to be very strong, due to the large
angular extent of the LSC, it should contribute to the anisotropy on
the on the angular scales of $\sim 1-10^{\circ}$.  The thermal SZ map
has patchy appearance similar to that of the X-ray brightness in
Figure~\ref{fig:sky2} with $\Delta T/T\sim 5\times 10^{-6}-10^{-5}$
within groups and clusters, $\sim 10^{-7}$ on their outskirts, and
$\sim 10^{-8}$ in the strongest filaments. The kinetic SZ effect has
similar magnitude. These fluctuations are small and are below current
sensitivity limits of the SZ observations; their contribution to the
large angular scale anisotropies measured by COBE satellite is
insignificant.

Finally, we presented (Figure~\ref{fig:sky2}) sky maps of column
densities of the three ionic species of oxygen: OVI, OVII, and OVIII.
The column densities were calculated using the densities and
temperatures of gas in the high-resolution region of the simulation
and assuming the uniform metallicity of $0.3$ solar and the observed
X-ray background.  Although, the OVI is the least abundant (its
abundance is more than an order of magnitude smaller than abundances
of OVII and OVIII) of the three ions, absorption or emission in its
resonant doublet at $\lambda\lambda 1031.92,\ 1037.62\ {\rm\AA}$
offers {\em currently\/} the best prospect for detection of the
warm/hot intergalactic gas in the LSC region. For the assumed
metallicity of the IGM, our simulations predict that the probability
to have a column density of OVI detectable with FUSE along a random
line of sight in the northern galactic hemisphere is $\gtrsim 0.5$;
the probability is higher in the regions of the sky close to the
supergalactic plane.

Our predictions would be affected if metallicity in filaments is
significantly smaller than 0.3 solar (the metallicities observed in
the virialized regions of clusters).  However, there are good reasons
to believe that the filaments should be enriched to the levels
comparable to the intracluster medium. The observations of metals in
the Lyman alpha forest at high redshifts indicates that regions with
overdensities of $\sim 10$ (corresponding to filaments) have
metallicities in the range $10^{-3}-10^{-2}$ solar \citep[see,
e.g.,][]{cowie_songaila98,schaye_etal00}. The simulations predict a
steep dependence of metallicity on the density in this regime with
metallicity quickly reaching values of $\sim 0.1$ solar in the
vicinity of the virialized regions
\citep[e.g.,][]{gnedin98,co99_chemical,aguirre_etal01}. As the
Universe evolves, the metallicity in the low-density regions is
expected to increase steadily \citep{co99_chemical}. At $z=0$,
numerical simulations of \citet{co99_chemical} predict the average
metallicity of $\sim 0.05-0.1$ solar (with a significant scatter
around the average) for overdensities of $\sim 10$; their results can
be approximated by the density-dependent metallicity given by
equation~\ref{eq:co}.  Figure~\ref{fig:skyfrac} shows that our results
are sensitive not to the particular density dependence of metallicity
but to the average metallicity of gas at overdensities $\approx 1-10$.
Detections (or lack thereof) of oxygen absorption within the Local
Supercluster should therefore provide useful constraints on the $z=0$
metallicity of IGM in filaments.

Although the structures in the simulation reproduce the large-scale
density field rather well, we expect certain differences with the real
observed structures. For example, the simulated Virgo cluster is
displaced by several degrees from the location of the real Virgo. We
cannot thus reliably predict the column density in any specific
direction. However, viewed statistically, our results suggest that for
a sample of background sources distributed over the northern sky,
about half of the lines of sight should pass through the column
density $\gtrsim 10^{14}{\ \rm cm^{-2}}$.  Unfortunately, the
probability decreases steeply for higher column densities ($\approx
10\%$ for $N_{\rm OVI}\gtrsim 4\times 10^{14}{\ \rm cm^{-2}}$. We
predict that OVI column density is very low in the area of $\approx
10^{\circ}$ around the Virgo cluster. This prediction is only for the
IGM; the halos of individual galaxies in or around Virgo may cause OVI
absorption.  The OVI absorption of the intergalactic gas should be
easy to separate from the local absorption in the Milky Way: most of
the absorbing gas should be redshifted by $\sim 3\pm 2\times 10^{-3}$.
The typical absorption should also be fairly broad as the typical line
of sight peculiar velocities of gas in the high density regions are
$\sim 100-400\ {\rm km\ s^{-1}}$ (see Figs.~5 and 6 in Paper I).

The absorption lines of OVII and OVIII in soft X-rays will be
difficult to detect with current instruments. The predicted
probability for absorption detectable with {\sl Chandra\/} or {\sl
XMM-Newton\/} is $\lesssim 5\%$ for a random line of sight in the
northern galactic sky. For the planned {\sl
Constellation-X}\footnote{{\tt http://constellation.gsfc.nasa.gov/}}
mission the probability is an order of magnitude higher.  The
simulations predict that regions of the highest column densities of
OVII and OVIII correspond to the high-density regions in and around
galaxy groups (see Fig.~\ref{fig:sky2}).

We would like to point out that our predictions are made under the
assumptions negligible cooling and absence of significant heating of
the intergalactic gas. We think that effects of cooling should not
have a significant effect on our predictions for the oxygen ion column
densities. The cooling time is sufficiently short only in the high
density regions which would occupy a very small fraction of the sky
and thus would not affect the cumulative distribution of column
densities. Comparison of adiabatic simulation to the simulations that
include cooling and stellar feedback presented by \citep{dave_etal01}
shows that the fraction of baryons in the warm/hot intergalactic
medium (that causes most of the absorption) is not sensitive to the
inclusion of additional physics. Note, however, that the effects of
feedback in these simulations are much smaller than in the extreme
feedback models that assume a uniform preheating of the intergalactic
gas with energy of order of 1~keV per particle in order to explain the
observed correlation of cluster X-ray luminosities and temperatures.
Such feedback would affect the gas evolution and its spatial
distribution lowering gas density in groups and poor clusters and
completely destroying most of the filaments. However, we believe that
there is no compelling evidence that the uniform preheating actually
occured.  Moreover, the existence of lyman alpha forest at high and
low redshifts argues against such extreme uniform feedback.

The results presented in this paper are only one example of useful
application of the constrained simulations. In Paper I we addressed
the problem of the coldness of the observed local peculiar velocity
field of galaxies and showed that the properties of the local flow can
be well reproduced in the currently popular {$\Lambda$}CDM cosmology.
The constrained simulations are unique because they allow properties
of individual simulated structures to be compared to the properties of
their real counterpart.  As a first step in this direction, we are
currently carrying out high-resolution constrained simulations of the
Virgo cluster in the {$\Lambda$}CDM cosmology.  The simulated cluster
forms in the environment very similar to the real Virgo. The
differences in the properties and internal structure of the simulated
and observed clusters should thus give us clues about the physical
processes shaping the evolution of intracluster gas.  Another very
interesting application of the constrained $N$-body$+$gasdynamics
simulations would be predictions on the spatial distribution of the
local lyman alpha forest. These predictions could be compared to and
provide interpretation for the observed large-scale distribution of
the nearby lyman alpha absorbers \citep[e.g.][]{stocke_etal00}. We
conclude that the constrained simulations of the type presented here
should have many other useful applications in comparisons of model
predictions with various observations of the nearby Universe.

\acknowledgements

We would like to thank Alexei Khokhlov for providing us with his
Godunov solver and very useful discussions during the initial stages
of code development and David Weinberg for useful discussions and
suggestions during the course of this project.  We acknowledge support
from the grants NAG-5-3842 and NST-9802787.  A.V.K.  was supported
by NASA through Hubble Fellowship grant from the Space Telescope
Science Institute, which is operated by the Association of
Universities for Research in Astronomy, Inc., under NASA contract
NAS5-26555. Y.H. has been supported in part by the Israel Science
Foundation (103/98).  AAK and AVK thank the Institute of Astronomy at
Cambridge in the spring of 2001 for hospitality and support during
their visit where the major portion of this project was completed.
Computer simulations presented in this paper were done on the Origin
2000 at the National Center for Supercomputing Applications (NCSA) at
Urbana-Champaign.

\bibliographystyle{apj}

\bibliography{lsc2}

\end{document}